\documentclass[aps,prl,twocolumn,showpacs,superscriptaddress]{revtex4}
\usepackage{amsmath}
\usepackage{dcolumn}
\usepackage{epsfig}
\usepackage{graphicx}
\usepackage{latexsym}

\def\BSCCO{Bi$_2$Sr$_2$CaCu$_2$O$_{8+\delta}$}

\begin{document}
\hyphenation{Ka-pi-tul-nik}



\title{Inhomogeneity Induces Resonance Coherence Peaks in Superconducting 
$\rm Bi_2Sr_2CaCu_2O_{8+\delta}$}

\author{A.~C.~Fang}
\affiliation{Department of Applied Physics, Stanford University, Stanford, CA 94305}
\author{L.~Capriotti}
\affiliation{Credit Suisse First Boston, Ltd. (Europe), One Cabot Square,
London E14 4QJ, UK}
\author{ D.~J.~Scalapino} 
\affiliation{Department of Physics, University of California, Santa Barbara, CA 93106-9530}
\author{S.~A.~Kivelson}
\affiliation{Department of Physics, Stanford University, Stanford, CA 94305}
\author{N.~Kaneko}
\affiliation{National Institute of Advanced Industrial Science and
Technology, Tsukuba Central 2-2, Tsukuba, Ibaraki 305-8568, Japan}
\author{M.~Greven}
\affiliation{Department of Physics, Stanford University, Stanford, CA 94305}
\author{A.~Kapitulnik}
\affiliation{Department of Applied Physics, Stanford University, Stanford, CA 94305} 
\affiliation{Department of Physics, Stanford University,
Stanford, CA 94305}

\date{\today}

\begin{abstract}
In this paper we analyze, using scanning tunneling spectroscopy,
the density of electronic states in nearly
optimally doped $\rm Bi_2Sr_2CaCu_2O_{8+\delta}$ in zero field.  Focusing on the superconducting gap, we find patches of what appear to be two different phases in a background of some average gap, one with a relatively small gap and sharp large coherence peaks and one characterized by a large gap with broad weak coherence peaks. We compare these spectra with calculations of the local density of states for a simple phenomenological model in which a $2\xi_0 \times 2\xi_0$ patch with an enhanced or supressed d-wave gap amplitude is embedded in a region with a uniform average d-wave gap.
\end{abstract}

\pacs{74.72.Hs, 74.50.+r, 74.25.-q}

\maketitle

\begin{figure}[h]
\begin{center}
\includegraphics[width=0.95 \columnwidth]{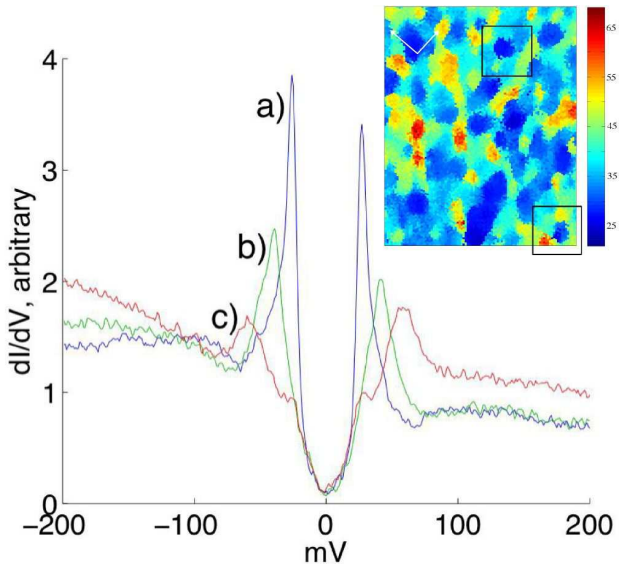}
\end{center}
\vspace{-4mm}
\caption{ Example spectra in a a) small, b) average, and c) large gap region. Inset is a  typical $220\AA \times 280\AA$ map of gap size.  Crystal axes and squares that relate to Fig.~\ref{cuts} are marked. } 
\label{indspec}
\end{figure}

One of the surprising features revealed by 
Scanning Tunneling Microscopy (STM) studies of the high T$_c$ superconductor, 
\BSCCO (BSCCO), is a pattern of patches of what appears to be  two different phases, 
with significant differences in their electronic structures  
\cite{cren,howald1,pan1,lang,fang1,ak1}.
There are regions of relatively small local gap, $\Delta(\vec r) \sim 25 - 35$meV in which the 
peak in the local density of states (LDOS) 
at $V=\Delta$ 
is relatively sharp in energy and 
the peak height is  very large.  Other regions have a larger gap, with $\Delta(\vec r) \sim 50 - 
75$meV and broad and small peaks (See Fig.~\ref{indspec}).  
It is tempting (as is widely assumed) to associate these very different electronic 
structures with two different bulk electronic phases:  the small gap regions, because they appear 
to have distinct coherence peaks, are identified as regions of ``good" superconductivity, whereas the large gap regions are like a pseudo-gap phase which competes with superconductivity.  This latter identification 
has found support from data suggesting that there is  a subtle form of local charge-density wave 
order with period near four lattice constants (``stripes'' or ``checkerboards'')  
\cite{howald2,fang1,ak1,vershinin,mcelroy2,oda1,rmp} which is most apparent in the large-gap regions 
\cite{mcelroy2,fang1}.  However, because the characteristic size of the regions ($L \simeq 
30\AA$) is not much larger than the superconducting coherence length (  $\xi_0\sim 15\AA$), 
it is clear that whatever the bulk character of each region, superconducting correlations can 
leak from one region into the other via the proximity effect \cite{howald1}, thus complicating any such 
identification.

In this paper we report results of STM studies on nearly optimally doped {\BSCCO} \cite{samples} with high spatial and fine energy resolution.  From this improved data we observe, as illustrated  in Fig.~\ref{indspec}, that:   {\bf 1)}  In the small gap regions, the peaks in the LDOS are 
too large to be the coherence peaks of a uniform BCS d-wave superconductor (see Fig.~\ref{indspec}a);
there is excessive spectral weight compared to the number of states pushed up from below the gap.
{\bf 2)} The peaks in the large gap regions are too broad and small to be the coherence peaks of a uniform 
BCS d-wave superconductor (see Fig.~\ref{indspec}c).
{\bf 3)}  These regions are interspersed in a background ``average" gap 
$\bar{\Delta}\equiv \Delta_0\approx 40$meV that produces a visible feature (typically, a shoulder) in the LDOS in 
 nearby regions;   this coincides with the gap inferred from angle resolved photoemission 
measurements \cite{feng1,valla}.  
  
To interpret these results, we have calculated the quasiparticle LDOS  for a mean field d-wave 
BCS model in which the strength of the pairing field (gap amplitude) is changed in a small $L 
\times L$ patch with $L \sim 2\xi_0$. We find that structure like that seen in the 
``small gap regions''  arises 
from resonant bound states if the gap amplitude 
vanishes (or is, at least, small compared to the peak energy) in the 
$L\times L$ patch. (see Fig.~\ref{fig3}b.)  Structure similar to that seen 
in the large gap regions is found if  a large pairing field is assumed inside such a patch 
\cite{nunner}.  (see Fig.~\ref{fig3}c.)  One is thus led to conclude that small gap regions with ``large coherence peaks'' are 
regions with a much smaller than average pairing potential. Conversely, the fact that the concentration of large gap regions increases in increasingly  underdoped samples, suggests that these regions, despite their strong pairing tendencies, have little or no superfluid density (phase stiffness) \cite{nature}. Finally, the fact that we see spectral features ``leaking" between regions
 suggests that we are seeing patches of  proximity coupled phases.

For tunneling perpendicular to the Cu-O planes,  a typical d-wave  BCS shape of the 
spectrum is expected, characterized by a ``v-shaped'' LDOS at low bias and  coherence peaks that 
accommodate the spectral weight  from the opening of a gap with nodes.   Early on, the general d-wave shape of STM spectra was confirmed in BSSCO \cite{renner1};  however data always appeared with significant particle-hole asymmetry in the background and subsequent analyses revealed  that very few spectra quantitatively fit  a BCS d-wave prediction, especially  the coherence peak strength and shape. 

Figure \ref{indspec} shows spectra often seen in the small, 
average, and large gap regions. In our analysis, we define the (positive bias) peak energy in the LDOS as the local gap $\Delta(\vec r)$.
In our samples, we find the average gap is $\Delta_0 \sim 40$ meV \cite{howald1,howald2,fang1,ak1} and that approximately 75$\%$ of the area has a gap that is within $\pm$10 meV of the average.
Embedded in this background  are 
patches of smaller gap ($\Delta \leq$ 30 meV) which cover approximately 15$\%$ of the area and 
of larger ($\Delta \geq$ 50 meV) which account for the remaining 10$\%$. 
In addition to the earlier comments regarding the coherence peak shape in these regions, 
we note that although the differences between the different spectra are more subtle\cite{howald1,lang} at energies far below the gap, the minimum tends to be more ``v-shaped'' in the large and average gap regions, and more rounded where the coherence peaks are anomalously large. 

\begin{figure}[h]
\begin{center}
\includegraphics[width=0.95 \columnwidth]{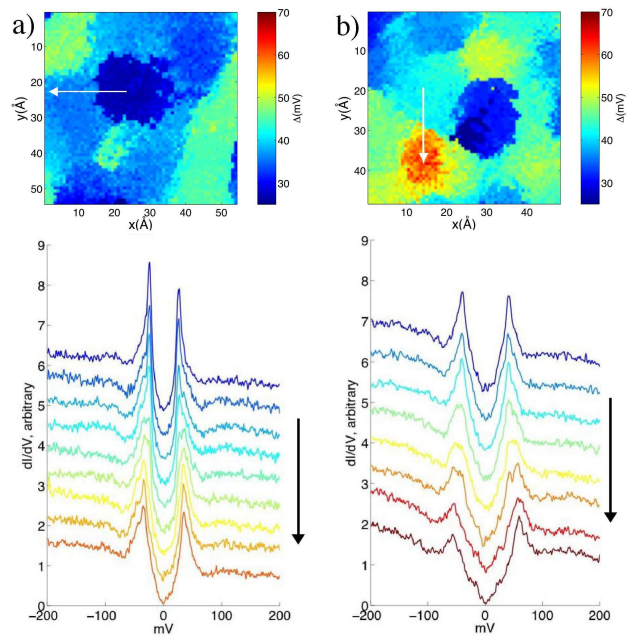}
\end{center}
\vspace{-4mm}
\caption{ a) Gap distribution and the  evolution of spectra as this region is crossed near a)  small gap region, and b)  large gap region. (Spectra follow arrows).} 
\label{cuts}
\end{figure}

\begin{figure}[h]
\begin{center}
\includegraphics[width=0.95 \columnwidth]{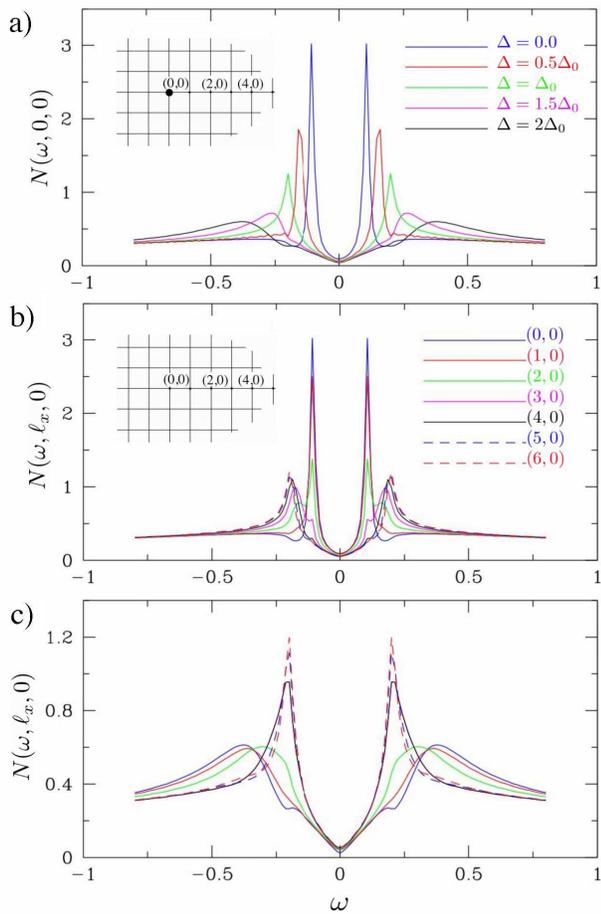}
\end{center}
\vspace{-4mm}
\caption{LDOS $N(\omega, \ell_x,\ell_y)$ versus $\omega$ for a $5\times 5$ patch centered at (0,0).
 a)  LDOS at the center of the cluster $N(\omega, 0, 0)$ versus $\omega$ for different
values of $\Delta (\ell)$.  b) $\Delta (\ell) = 0$ on the patch.  c) $\Delta (\ell)=2\Delta_0$ on the patch and for both, $N(\omega, \ell_x,\ell_y)$ is shown for sites along $(\ell_x,0)$. }\label{fig3}
\end{figure}

To study the behavior of the LDOS spectra as we go from a region of one gap size to another, we 
initially take a scan over a large area. Then we select several small areas
and study them in detail, with a resolution of several spectra per atom.  
To maximize energy resolution, we limited the bias modulation used to acquire the $dI/dV$ data 
to 2mV, and applied minimal filtering for the data collection. Figure \ref{cuts} shows maps of $\Delta(\vec r)$ with a small (a) and large (b) gap region. Below each figure we also show line cuts of spectra along the arrow.  

The spectra in Fig.~\ref{cuts}a illustrate the evolution of the LDOS on going from a small gap 
region with $\Delta < 30$ meV, to an ``average" gap region with $\Delta_0 \sim 40$ meV.  
The crossover from one type of spectrum to the next occurs over a  distance  $\leq \xi_0$ with the anomalously large ``coherence peaks" diminishing in strength while
 new peaks at the background gap of that region (typically  $\sim \Delta_0$) increase in intensity.
The signature at $\Delta_0$ can be followed throughout the spectra from top (blue) to bottom (red):  Deep in the small gap region it  appears as a weak shoulder above the main peak.  At the border, a 
two-peak structure is apparent - one corresponding to the small gap characteristics  
and the other near $\Delta_0$.  Finally, outside the small gap region, the main peak occurs at about 
$\Delta_0$ and has a more BCS-like structure.  Line cuts through other small gap regions show similar behavior, although in instances of small gap differences, the two-peak structure is harder to see. 

 Figure \ref{cuts}b shows a line cut that starts in the average gap region, and ends up at the 
 center of a large gap region. Again, the peak corresponding to 
the average gap diminishes in strength without dispersing in energy, while a new broad peak appears at $\Delta \sim 65$ meV  and the ``average gap"  coherence peak becomes a shoulder inside the large gap.  Other line cuts in different large gap regions of the sample show similar evolution of the features, 
 with the shoulder inside the gap being more/less visible when the differences 
 in $\Delta$ are large/small.
 
Finally, as can be noted from Fig.~\ref{indspec} (inset), while small gap regions may appear isolated in the background of the average gap, large gap regions almost always appear within a distance $< \xi_0$  of small gap regions.  This feature  may reflect the optimal-doping samples we are using which favor the creation of small gap regions. 
 
 To obtain insight into the implications of these results, we have computed the LDOS for a simple model Hamiltonian, meant to represent an effective mean-field Hamiltonian for the quasiparticles:

\begin{equation}
 \mathcal{H} =  \sum_{\ell,\delta}\left[-t\psi_\ell^\dagger \tau_3 \psi_{\ell+\delta}
+ (-)^\delta \frac{\Delta(\ell)}{8}\psi_\ell^\dagger \tau_1 
\psi_{\ell+\delta}+ h.c.\right]
 \end{equation}

\noindent where the vectors $\ell$ label the lattice sites,  $\delta$ are the nearest-neighbor vectors, and
$(-)^\delta = 1$ for $\delta= \pm \hat{x}$ and $(-)^\delta=-1$ for $\delta = \pm \hat{y}$.  
In these expressions, we have adopted the usual Nambu notation with 
$\psi_\ell^\dagger = (c_{\ell \uparrow}^\dagger , c_{\ell \downarrow})$.  
In the uniform case, $\Delta(\ell)=\Delta_0$, $\mathcal{H}$ 
describes a uniform square lattice with near-neighbor hopping $t$ and a d-wave mean field 
characterized by a gap $\Delta(k) = \frac{\Delta_0}{2}(\cos~k_x - \cos~k_y)$. 
In the following we will consider 
the non-uniform situation in which $\Delta(\ell)=\Delta$ on the sites inside an 
 $L \times L$ cluster embedded in a much larger ($M \times M$ with $M \gg L$) cluster in which $\Delta(\ell)=\Delta_0$.   All the calculations shown in the present paper are for $M=800$ and $L=5$, although we have performed calculations for a range of $M$'s and confirmed that $M=800$ is large enough that the results are independent of $M$.  

We are interested in determining how the LDOS $N(\omega, \ell)$ varies as one 
moves from outside the cluster to sites inside the cluster, where

\begin{eqnarray}
\label{dos}
N(\omega, \ell) =&& N(\omega)\\
-&&\frac{1}{\pi} \mathcal{I}m\left[{\sum_{\ell_1,\ell_2}} ^\prime
Tr\left( G(\ell-\ell_2)T(\ell_2,\ell_1)G(\ell_1-\ell)\right)\right] 
\nonumber
\end{eqnarray}

\noindent Here $N(\omega)$  is the average density of states, $\sum^\prime$ runs over sites inside the $L\times L$ patch, $G(\ell)$ is the single particle Green's function of the uniform lattice

\begin{equation}
G(\ell) = \frac{1}{N}\sum_k \left(\frac{\omega + \epsilon_k \tau_3 + \Delta_k \tau_1}{\omega^2 - 
\epsilon_k^2 - \Delta_k^2}\right)e^{ik\cdot \ell}
\label{green}
\end{equation}

\noindent and $T(\ell_2, \ell_1)$ is the $T$-matrix associated with the scattering ``potential''
$\tilde\Delta(\ell)\equiv \Delta(\ell)-\Delta_0$,

\begin{eqnarray}
\nonumber  T(\ell_2,\ell_1) = \sum_\delta (-)^\delta \frac{\tilde\Delta(\ell_2)}{8}\tau_1 
\delta_{\ell_2,\ell_1 -\delta}  \\   + {\sum_{\ell_3,\delta}}^\prime(-)^{\delta}
\frac{\tilde\Delta(\ell_2)}{8}\tau_1 G(\ell_2+\delta -\ell_3)T(\ell_3,\ell_1)\ .
 \label{tterm}
\end{eqnarray}

Representative results of our calculations are shown in Fig.~3 for $\Delta_0/t = 0.2$ and a 
small damping factor of 0.01.  Results for the density of states at the center of a $5\times 5$
cluster in which the gap amplitude in the cluster ranges from $\Delta=2\Delta_0$ to $\Delta=0$ are shown in Fig.~3a. Here, one sees that when the gap amplitude in the cluster is large compared to the background, the density of states at the center of the patch has a broad response at $\omega = \pm \Delta$.  However, as the cluster gap amplitude decreases below $\Delta_0$, resonant peaks develop below $\pm \Delta_0$ and move down in energy as  $\Delta$ decreases.
The height of the resonance peaks also increases as 
$\Delta$ decreases.  In order to illustrate the spatial dependence of $N(\omega, \ell)$, we consider the case in which $\Delta=0$, corresponding to a zero pairing amplitude in the $5\times 5$ patch.  For
this case, the density of states $N(\omega, \ell)$ versus $\omega$ for various sites $(\ell_x,
\ell_y)$ are shown in Fig.~3b. Here, the site (0, 0) corresponds to the center of the 
cluster and results are shown for $\ell_y=0$ with $\ell_x$ varying from 0 to 6. 
For sites inside the ``gapless" cluster 
one sees a resonant response. This response appears at a lower energy than $\Delta_0$ and
the peak height, for a given broadening, is significantly larger than the logarithmic structure 
in the bulk d-wave density of states.
As one moves out from the center of the cluster, the sharp low energy peak in the LDOS loses intensity and weight begins to grow at $\omega\sim\Delta_0$, so that at the boundary of the cluster a two-peak structure is clearly observed.   Outside the cluster, the LDOS returns to its average behavior within a few lattice constants.

If the gap parameter is doubled inside the cluster, $\Delta=2\Delta_0$, one has $N(\omega,\ell)$ shown in  Fig.~\ref{fig3}c.  In this case, for sites inside the cluster, 
$N(\omega,\ell)$ exhibits a broadened response near $2\Delta_0$ as well as a weak response at 
$\Delta_0$.  Here again, as one moves several lattice spacings outside the cluster, 
the density of states returns to its uniform behavior, characterized by the logarithmic coherence 
peaks at $\omega = \pm \Delta_0$.   
Note the change in scale between 
Figs.~\ref{fig3}a,b and \ref{fig3}c and how much stronger the resonance peaks are 
compared to the logarithmic  peaks.

The low energy behavior of $N(\omega,\ell)$ is 
less dramatically $\ell$ dependent than the peak structure.  None-the-less, we believe that it is significant that the gap minimum is more ``v-shaped'' near the center of the large-gap cluster, and more rounded near the center of the small-gap cluster.
We have not systematically explored the dependence of the results on the size of the cluster, $L$, but we have checked that similar behavior is obtained for somewhat different sized clusters and for clusters rotated by 45$^o$. 

The model we have solved is admittedly overly simple, especially in that it neglects the strong
on-site Hubbard $U$.
Nonetheless, the  qualitative similarities between features of the model calculations and the STM 
data suggest that some aspects of the problem are being 
successfully modeled.  
Clearly there are structural 
variations from place to place in {\BSCCO};  for instance,  a positive correlation between the 
concentration of Oxygen interstitials and the large gap regions have been reported by  McElroy 
{\it et al.} \cite{mcelroy2}.  However, can structural differences due to 
these impurities 
give rise to a significant enhancement of the local pairing amplitude and what about the 
regimes in which the gap is suppressed?  

Rather, it would seem that some type of intrinsic amplification 
of the effect of the structural variations is likely to be essential.  For instance, if (as has been previously suggested \cite{firstorder})  doped antiferromagnets are near the cusp 
of a first order transition between two electronically distinct states, then small differences in 
the local structure can nucleate small regions of one phase or the other.   It has 
been well established by now that above optimal doping ($x\approx $ 0.15 holes per Cu atom in the 
Cu-O plane) the average gap decreases with increasing doping in proportion to $T_c$, roughly as 
$2\Delta \approx 8k_BT_c$.  In underdoped samples, the average gap increases with decreasing 
doping, rising from $\Delta_0 \sim 40$meV  at optimal doping to $\Delta_0 \sim 55$meV at around 
$x\approx 0.05$, where $T_c \to 0$ \cite{shenrmp}.  Correspondingly, STM studies of underdoped 
{\BSCCO} reveal that the fraction of large gap regions increases with decreasing 
$x$, and the fraction of small gap regions decreases \cite{cren,howald1,lang}.  Thus, it would be 
natural to identify the large gap regions as being more representative of the electronic 
structure of underdoped cuprates and the small gap regions more representative of overdoped 
cuprates, both influenced by some ``average gap" background. However, decreasing $x$ also leads to a rapid decrease of $T_c$ and the superfluid density, which implies that a large pairing field alone is
insufficient to characterize the features of the electronic structure which reflect the approach to the Mott insulator.

\acknowledgments

ACF and AK acknowledge support by DoE grant DE-FG03-01ER45925. LC acknowledges support of NSF grant Phy99-07949 at KITP (UCSB) where most of the reported calculations  were carried out when he was a KITP post doctoral fellow. LC and DJS acknowledge useful discussions with L. Balents and support under NSF Grant No.~DMR02-11166. Crystal growth was supported by DoE grants No.~DE-FG03-99ER45773 and No.~DE-AC03-76SF00515.


\begin{thebibliography}{99}

\bibitem{cren}
T.~Cren {\it et al.}, {\sl Phys.~Rev.~Lett.} {\bf 84}, 147 (2000).

\bibitem{howald1}
C.~Howald, P.~Fournier, and A.~Kapitulnik, {\sl Phys.~Rev.~B} {\bf 64},
100504 (2001).

\bibitem{pan1}
S.H.~Pan {\it et al.}, {\sl Nature} {\bf 413}, 282
(2001).

\bibitem{lang}
K.M.~Lang {\it et al.}, {\sl Nature} {\bf 415}, 412 (2002).

\bibitem{fang1}
A.~Fang {\it et al.}, {\sl Phys.~Rev.~B} {\bf 70}, 214514 (2004).

\bibitem{ak1}
A.~Kapitulnik, A.~Fang, C.~Howald, and M.~Greven, cond-mat/0407743,  
accepted for publication in {\sl J.~Phys.~Chem.~Solids}, SNS2004 Special issue.

\bibitem{oda1}
N.~Momono, A.~Hashimoto, Y.~Kobatake, M.~Oda, M.~Ido; cond-mat/0505254 (2005).

\bibitem{rmp}
S. A. Kivelson {\it et al.}, \rmp {\bf 75}, 1201 (2003).

\bibitem{howald2}
C.~Howald, H.~Eisaki, N.~Kaneko, and A.~Kapitulnik; cond-mat/0201546, 
{\sl Proc.~Natl.~Ac.~Sci.} {\bf 100}, 9705 (2003);  C.~Howald {\it et al.}, {\sl Phys.~Rev.~B} {\bf 67}, 014533 (2003).

\bibitem{mcelroy2}
K.~McElroy {\it et al.}, {\sl Phys.~Rev.~Lett.} {\bf 94}, 197005 (2005).

\bibitem{vershinin}
M.~Vershinin {\it et al.}, {\sl Science} {\bf 33}, 
1995 (2004).

 \bibitem{samples}
Samples are prepared as slightly overdoped with $T_c \approx$ 89 K. Cleaving at room temperature at UHV conditions cause some oxygen to leave the exposed Bi-O surface layer yielding  
samples  closer to optimal doping \cite{howald1,oda2}.
 
 \bibitem{oda2}
M.~Oda, C.~Manabe, and M.~Ido, {\sl Phys.~Rev.~B} {\bf 53}, 2253 (1996).

\bibitem{feng1}
D.L.~Feng {\it et al.}, {\sl Science} {\bf 289}, 277 (2000).

\bibitem{valla}
T.~Valla {\it et al.}, {\sl Science} {\bf 285}, 2110
(1999).

\bibitem{nunner}
Similar effects are seen in a self-consistent Bogolubiv-deGennes calculation for a model with a random dopant-modulated pairing interaction in: T.S.~Nunner, B.M.~Andersen, A.~Melikyan, and P.J.~Hirschfeld; cond-mat/0504693 (2005).

\bibitem{nature}  V. J. Emery and S. A. Kivelson, {\it Nature} {\bf 374}, 434 (1995).


\bibitem{renner1}
C.~Renner {\it et al.}, {\sl Physica B} {\bf 194}, 1689 (1994).

\bibitem{firstorder}
V.J.~Emery, S.A.~Kivelson, and H.Q.~Lin, {\sl Phys.~Rev.~Lett.} {\bf 64}, 475 (1990); 
S. A. Kivelson, G. Aeppli, and V. J.  Emery,  {\sl Proc.~Natl.~Ac.~Sci.}
{\bf 98}, 11903 (2001).

\bibitem{shenrmp}
See {\it e.g.} the review by A.~Damascelli, Z.~Hussain, and Z-X.~Shen, 
{\sl Rev.~Mod.~Phys.} {\bf 75}, 473 (2003).


\end{thebibliography}
\end{document}